\documentclass[journal=jacsat,manuscript=article]{achemso}

\usepackage[version=3]{mhchem} 
\usepackage{siunitx}
\usepackage[capitalize]{cleveref}
\usepackage{float}

\author{Thomas Descamps}
\email{descamps@kth.se}
\affiliation{Department of Applied Physics, KTH Royal Institute of Technology, Roslagstullsbacken 21, 10691 Stockholm, Sweden}
\author{Tanguy Schetelat}
\affiliation{Department of Applied Physics, KTH Royal Institute of Technology, Roslagstullsbacken 21, 10691 Stockholm, Sweden}
\author{Jun Gao}
 \affiliation{Department of Applied Physics, KTH Royal Institute of Technology, Roslagstullsbacken 21, 10691 Stockholm, Sweden}
 \author{Philip J. Poole}
 \affiliation{National Research Council Canada, Ottawa, Ontario K1A 0R6, Canada}
 \author{Dan Dalacu}
\affiliation{National Research Council Canada, Ottawa, Ontario K1A 0R6, Canada}
\author{Ali W. Elshaari}
 \affiliation{Department of Applied Physics, KTH Royal Institute of Technology, Roslagstullsbacken 21, 10691 Stockholm, Sweden}
\author{Val Zwiller}
\email{zwiller@kth.se}
\affiliation{Department of Applied Physics, KTH Royal Institute of Technology, Roslagstullsbacken 21, 10691 Stockholm, Sweden}
\alsoaffiliation{Single Quantum BV, Delft, The Netherlands}

\title{Acoustic modulation of individual nanowire quantum dots integrated into a hybrid thin-film lithium niobate photonic platform}

\begin{document}

\begin{abstract}
Surface acoustic waves (SAWs) are a powerful tool for controlling a wide range of quantum systems, particularly quantum dots (QDs) via their oscillating strain fields. The resulting energy modulation of these single photon sources can be harnessed to achieve spectral overlap between two QDs otherwise emitting at different wavelengths.
In this study, we integrate \ce{InAsP}/\ce{InP} nanowire quantum dots onto a thin-film lithium niobate platform, a strong piezoelectric material, and embed them within \ce{Si_3N_4}-loaded waveguides. We demonstrate emission wavelength modulation of $\SI{0.70}{nm}$ at $\SI{13}{dBm}$ with a single focused interdigital transducer (FIDT) operating at $\SI{400}{MHz}$, and achieve twice this modulation by using two FIDTs as an acoustic cavity. Additionally, we bring two QDs with an initial wavelength difference of $\SI{0.5}{nm}$ into resonance using SAWs. This scalable strain-tuning approach represents a significant step towards producing indistinguishable single photons from remote emitters heterogeneously integrated on a single photonic chip, and paves the way for large scale on-chip quantum information processing using photonic platforms.
\end{abstract}

\section{Keywords}
quantum dots, single photon source, surface acoustic waves, thin-film lithium niobate, integrated photonics

\section{Introduction}

Surface acoustic waves (SAWs), with their capacity to interact mechanically with both the supporting crystal and the materials on its surface, have shown significant interest for controlling various quantum systems, including superconducting qubits \cite{Gustafsson2014, Manenti2017, Moores2018}, spin qubits \cite{Jadot2021, Hermelin2011, McNeil2011}, quantum optomechanical cavities \cite{Balram2016, Jiang2020}, and single-photon emitters based on defect centers \cite{Golter2016, Lukin2020, Patel2022} or III/V semiconductor quantum dots (QDs). In the latter case, the oscillating electric field created by the SAW propagating on a piezoelectric medium was used to transport charge carriers to the QD and to control the emitter's charge state \cite{Hernandez-Minguez2012, Couto2009, Volk2012}. Additionally, the oscillating strain field induced by the SAW modulates the energy levels of the QD \cite{Weiss2018a}. Utilizing this property, coherent coupling between acoustic phonons and single photons \cite{DeCrescent2022, Weiss2021, Decrescent2024, Villa2017} as well as single-photon frequency shifting have been demonstrated \cite{Descamps2023, Vogele2020, Nysten2020}.
These investigations, predominantly focused on a single QD, could be extended to multiple emitters on the same chip, each independently modulated by a SAW to tune their emission wavelengths. This advancement would be of technological interest, as it would address the variance in emission wavelengths of these sources \cite{Tongbram2017, Rastelli2009, Paul2017, Laferriere2022}, a major limitation for their applications in integrated linear quantum computing \cite{OBrien2007, Kok2007, Moody2022, Dusanowski2023} and quantum communication \cite{Luo2023, Llewellyn2020}, where photon indistinguishability is paramount. Typically generated by driving an interdigital transducer (IDT) patterned on a piezoelectric substrate with a microwave signal, SAWs offer several advantages over other tuning mechanisms. First, the emission wavelength can be either redshifted or blueshifted, unlike thermo-optic schemes based on local heating of the source, which always result in a redshift \cite{Faraon2009, Elshaari2017}. Secondly, QDs can be directly modulated without the need for doping the heterostructure and making electrical contacts, as required for Stark effect-based tuning \cite{Schnauber2021, Ellis2018}. Lastly, the localized strain field and fabrication simplicity of this method make it more scalable and robust compared to other strain mechanisms, such as those using global static fields applied with piezoelectric substrates \cite{Elshaari2018, Yang2024} or MEMS technologies employing suspended films \cite{Quack2023, Grosso2017}.\\
In this work, we examine \ce{InAsP}/\ce{InP} nanowire (NW) quantum dots (NWQDs), which are known for being bright sources of high-purity and indistinguishable single photons \cite{Laferriere2022, Reimer2016, Yeung2023}. Unlike the monolithic approach, where self-assembled QDs are embedded in waveguides etched into the III/V heterostructure \cite{Buhler2022, Dusanowski2023}, the site-controlled NWs are picked up and placed \cite{Kim2017, Mnaymneh2020} onto an unreleased thin-film lithium niobate (LN) platform, as this strong piezoelectric material enables more efficient electro-mechanical transduction. The NWs are then integrated into \ce{Si_3N_4}-loaded waveguides \cite{Han2021, Han2022, Jiang2022}, and positioned at the center of an acoustic delay line consisting of focused interdigital transducers (FIDTs). We achieve a modulation of the emission wavelength of $\SI{0.70}{nm}$ by driving a single FIDT at $\SI{400}{MHz}$ with a microwave power of $\SI{13}{dBm}$, and we double this modulation by driving two FIDTs as an acoustic cavity. Finally, we demonstrate that two waveguide-integrated NWQDs, whose emission wavelengths differ by $\SI{0.5}{nm}$, can be brought into resonance using SAWs. This result paves the way for generating indistinguishable single photons from multiple remote QDs on a single heterogeneous photonic integrated chip.

\begin{figure}
\includegraphics[width=0.95\textwidth]{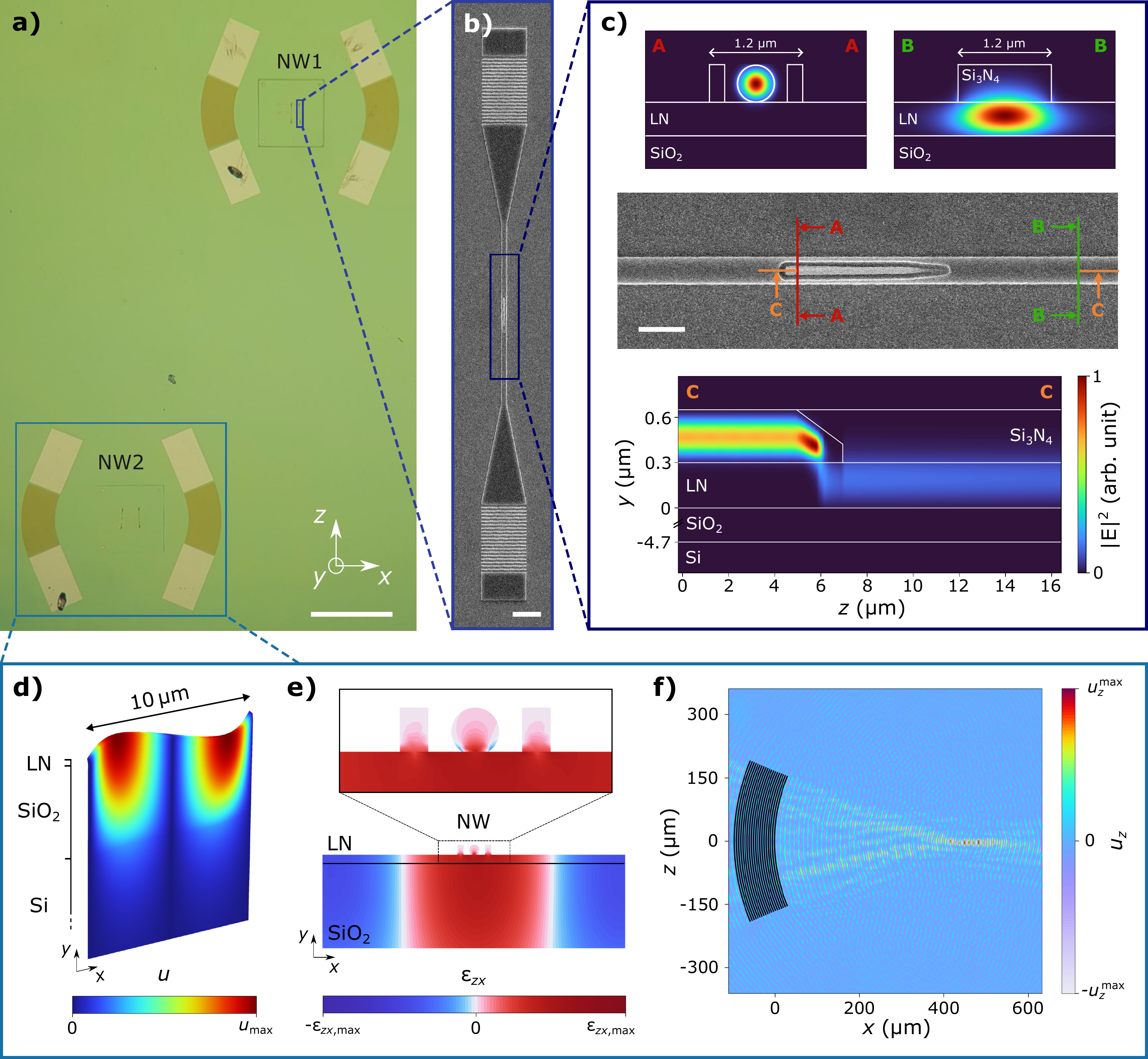}
\caption{\label{fig:Fig1}\textbf{a)} Optical microscope image of two acoustic delay lines designed for the independent acoustic modulation of two remote nanowires, each integrated into their own \ce{Si_3N_4}-loaded thin-film LN waveguides. The scale bar represents $\SI{500}{\micro m}$. \textbf{b)} Scanning electron microscope image of a nanowire within a waveguide from a similar device. The grating at the bottom is used for coupling photons from the waveguide to out-of-plane, and vice versa. The scale bar represents $\SI{5}{\micro m}$. \textbf{c)} Scanning electron microscope image of the integrated nanowire. The scale bar represents $\SI{2}{\micro m}$. Cross section AA shows the fundamental mode confined in the NW, while cross section BB displays the fundamental TE mode of the \ce{Si_3N_4}-loaded thin-film LN waveguide after the mode transfer region. Cross section CC illustrates the optical TE mode transfer from the NW to the strip-loaded waveguide. \textbf{d)} Displacement profile of the shear SAW mode SH0 obtained by COMSOL simulation. \textbf{e)} Strain profile generated by the SAW in the upper layers of the sample and in the NW placed on top. \textbf{f)} Displacement field of a SAW excited at $\SI{400}{MHz}$ by an FIDT with a $\SI{400}{\micro m}$ focal length and a $\SI{45}{^\circ}$ opening angle. The FIDT has a period of $\SI{10}{\micro m}$ with two electrodes per period.
}
\end{figure}

\section{Design and methods}
An optical microscope image of the hybrid quantum photonic platform developed in this work is shown in \cref{fig:Fig1}(a), featuring four nanowire quantum dots each integrated into a photonic waveguide and positioned within an acoustic delay line. The wurtzite \ce{InP} NWs embedding individual \ce{InAsP} QDs \cite{Dalacu2012, Laferriere2023} emitting around $\SI{900}{nm}$ were picked up with a nano-manipulator inside a scanning-electron microscope (SEM), and transferred to a $\SI{300}{nm}$-thick Y-cut thin-film LN chip with $\SI{4.7}{\micro m}$ buried \ce{SiO_2}. The NWs were oriented along the crystallographic Z-axis. A $\SI{350}{nm}$-thick \ce{Si_3N_4} loading layer was then deposited by plasma-enhanced chemical vapor deposition (PECVD) on the whole surface and etched to define the photonic elements. The waveguides were $\SI{1.2}{\micro m}$-wide and terminated with grating couplers used for exciting the QD and collecting the emitted photons. An SEM image of the photonic channel around the NW is presented in \cref{fig:Fig1}(b), while an SEM image of the waveguide-integrated NW is shown in \cref{fig:Fig1}(c). The alignment of the waveguide relative to the NW was well-achieved, with a $\SI{150}{nm}$-large gap present between them, as the \ce{Si_3N_4} did not reproducibly adhere to the \ce{InP} during deposition. The tapered shape of the NW favors an adiabatic mode transfer of the transverse electric (TE) mode of the NW to the fundamental TE mode of the waveguide. The latter is mainly confined in the LN since \ce{Si_3N_4} has a slightly lower refractive index. Finite-difference time-domain simulations (Lumerical) were conducted assuming lossless materials and yielded a coupling efficiency of $\SI{78}{\%}$.
Two of the four NWs, hereafter referred to as NW1 and NW2 (with quantum dots QD1 and QD2, respectively), were selected based on their emission properties to be at the center of two acoustic delay lines. Each delay line comprised two opposing FIDTs made of chromium with a common geometric focal point. Both FIDTs feature the same geometry, with a period of $ \Lambda = \SI{10}{\micro m}$ repeated $N=20$ times, a $\SI{400}{\micro m}$ focal length and a $\SI{45}{^\circ}$ opening. By orienting the transducers toward the X-axis of the crystal, a shear SAW with a fundamental frequency at $\nu_{0} = \SI{402.4}{MHz}$ can be excited. The displacement profile of this SH0 mode is shown in \cref{fig:Fig1}(d). Based on the delta-function model \cite{Hashimoto2012}, the bandwidth of the resonance is $\Delta \nu=\SI{17.8}{MHz}$, according to the expression $\Delta \nu = 2 \beta \nu_0 / (N \pi)$, with $\text{sinc}(\beta)=1/\sqrt{2}$. The in-plane displacement is perpendicular to the SAW propagation direction and is mostly confined into the LN and \ce{SiO_2} layers. The wave velocity is $c_{SH0} = \Lambda \times \nu_{0} = \SI{4024}{m \per s}$. The primary component of the associated strain tensor is the shear element $\varepsilon_{zx}$ whose profile is represented in \cref{fig:Fig1}(e). The presence of non-zero strain at the center of the nanowire, positioned on top of the thin-film LN, indicates that the QD experiences an oscillating strain field as the SAW propagates.
Compared to a straight-electrode IDT, which generates plane-wave SAWs, a focused IDT, whose electrodes are shaped as arcs of periodically spaced concentric circles, can be used to enhance the SAW intensity. The SAW radiated by the fabricated FIDT was simulated with COMSOL, and its transverse displacement field is displayed in \cref{fig:Fig1}(f). The maximum acoustic amplitude is reached at $x_0=\SI{470}{\micro m}$, offset by $\SI{70}{\micro m}$ from the geometric focal point of the FIDT. The acoustic intensity can be fitted to a Gaussian beam profile to extract a Rayleigh length of $x_\text{R}=\SI{60}{\micro m}$. Compared to a straight-electrode IDT, the acoustic field at the beamwaist is increased by a factor of 4.1, and at the geometric focal point by a factor of 2.7 (section S3 of the Supporting Information). The strain field experienced by the QD is therefore significantly enhanced due to the focusing capability of the FIDT.

\begin{figure}[ht]
\includegraphics[width=0.95\textwidth]{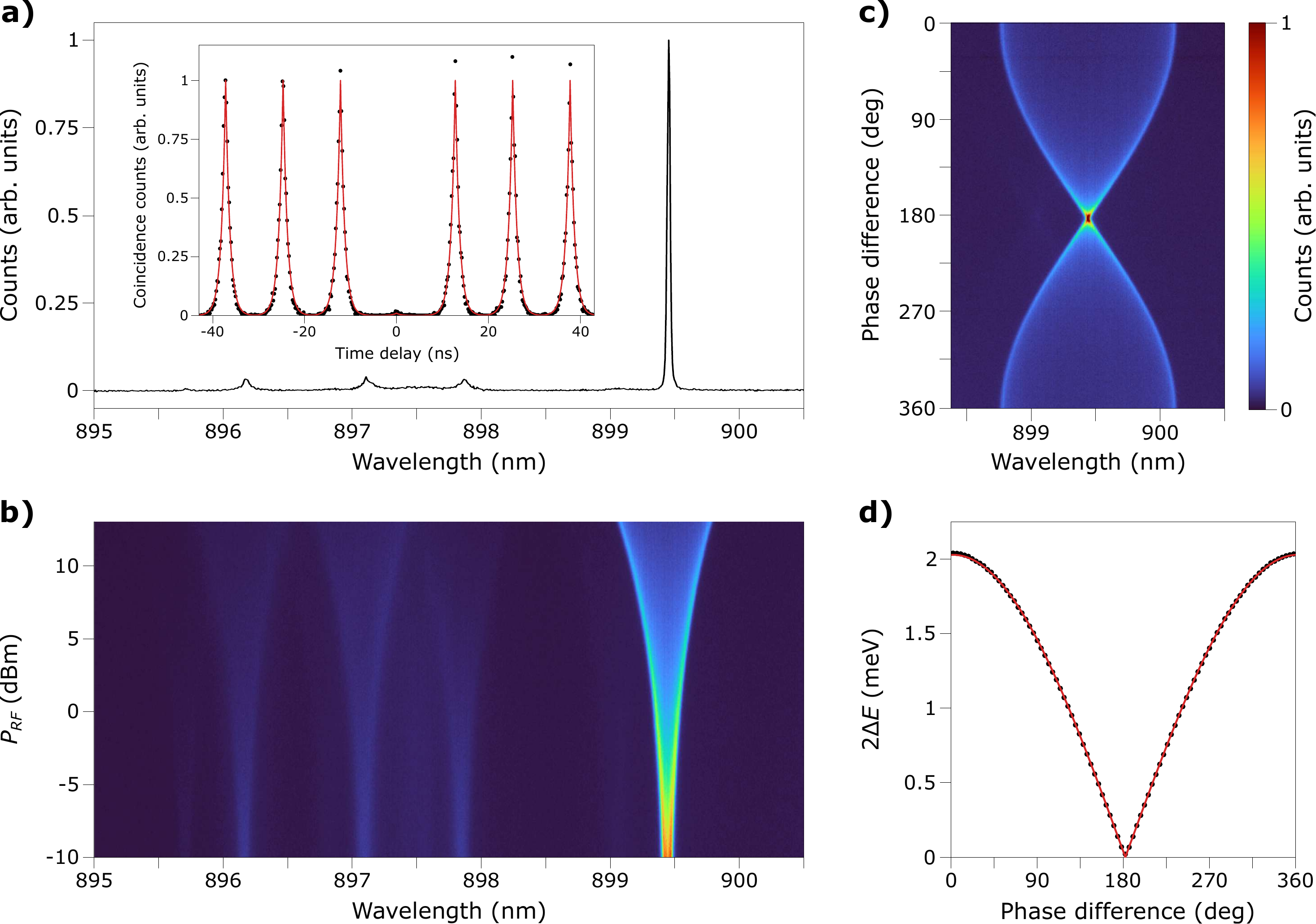}
\caption{\label{fig:Fig2}\textbf{a)} PL spectrum of QD1 without modulation. Inset, second order correlation function of the brightest emission peak. The fitting function is detailed in the main text. \textbf{b)} Optical modulation induced by a single SAW as a function of $P_{RF}$ (generation at $\SI{400}{MHz}$). The colorbar is the same as \textbf{c)}. \textbf{c)} Optical modulation of the brightest emission peak by two counter-propagating SAWs as a function of their relative phase $\Delta \phi$. Both FIDTs are excited at $\SI{400}{MHz}$ ($P_{RF}=\SI{12.5}{dBm}$). \textbf{d)} Strain-induced energy splitting as a function of $\Delta \phi$. The fitting function in red is detailed in the main text. For all the measurements, the QD was excited with a $\SI{80}{MHz}$ pulsed-laser at $\SI{500}{nW}$.}
\end{figure}

The sample was investigated at $\SI{1.8}{K}$ in a dry cryostat configured for confocal micro-photoluminescence (PL) measurements and equipped with high-frequency cables. An $\SI{80}{MHz}$ pulsed-laser (measured $\SI{80.026}{MHz}$) was focused with a microscope objective on one grating coupler to excite the waveguide-integrated NWQD above-band at \SI{800}{nm}. The PL signal propagating towards the same grating coupler was collected by the same microscope objective, dispersed by a $\SI{750}{cm}$ focal length spectrometer and detected by a liquid nitrogen-cooled charge-coupled device (CCD) camera. A two-channel analog signal generator was used to apply sinusoidal radio frequency (RF) signals with adjustable power $P_{RF}$ and phase difference $\Delta \phi$ to one or both FIDTs of the delay line.

\section{Results}
\cref{fig:Fig2}(a) displays the PL spectrum of QD1 without acoustic modulation at an excitation power of $\SI{500}{nW}$. In the following, we investigated the brightest line at $\SI{899.46}{nm}$, attributed to the charged exciton \cite{Laferriere2021}. After filtering with a monochromator ($\SI{0.1}{nm}$ bandwidth), the purity of the single photon source was assessed in a Hanbury Brown-Twiss measurement (inset of \cref{fig:Fig2}(a)). The signal was detected by superconducting nanowire single photon detectors and counted by a time tagging device. The second-order correlation function was fitted with a sequence of equidistant photon pulses assuming a mono-exponential decay, yielding a radiative decay time of $\tau = 0.88 \pm \SI{0.02}{ns}$. The suppression of the peak at zero time delay indicates strong single-photon emission. The ratio of the area of the zero time delay peak to the area of the finite time delay pulses gives $g^{(2)}(0) = 0.010 \pm 0.002$.
When a $\SI{400}{MHz}$ RF signal is applied to the FIDT, the sinusoidal modulation of the strain field around the QD induces a modulation of its bandgap energy at the same frequency, causing the spectral lines to oscillate around their unstrained energies \cite{Descamps2023}. Spectral detuning already becomes noticeable for all peaks at approximately $P_{RF}=\SI{-10}{dBm}$ and reaches $\SI{0.70}{nm}$ at $\SI{13}{dBm}$ (\cref{fig:Fig2}(b)). This optomechanical coupling arises exclusively from shear strain modulating the energy levels of the QD, an effect less commonly studied compared to normal strain coupling. Although the nanowire is not encapsulated inside the waveguide, it maintains good mechanical contact with the lithium niobate thin film even at moderate RF powers, as evidenced by the stable increase in modulation. The broadening also remains symmetric around the unstrained emission, indicating that heating of the QD is effectively mitigated at moderate RF powers \cite{Buhler2022}. By choosing a modulation frequency lower than the decay rate of the emitter, phonon sidebands around the central emission line are avoided.
Then, both FIDTs forming the delay line are driven at $\SI{400}{MHz}$ with two independent microwave channels to produce two counter-propagating SAWs whose superposition forms a standing wave. A minor performance discrepancy between the two FIDTs, attributed to fabrication imperfections, is compensated by applying slightly less power to the first FIDT ($P_{RF,1}=\SI{12.5}{dBm}$) compared to the second ($P_{RF,2}=\SI{13}{dBm}$). The standing wave generates a pattern of nodes (points of zero displacement) and anti-nodes (points of maximum displacement) whose position with respect to the nanowire can be adjusted by modifying the phase difference $\Delta \phi$ of the two RF signals. Figure 2(c) illustrates the modulation of the brightest emission line of QD1 as a function of $\Delta \phi$. When both signals are in phase, the nanowire lies at an anti-node of the standing wave, resulting in a modulation amplitude that is twice that obtained with a single propagating SAW. Conversely, the acoustic modulation is completely suppressed when a $\pi$ phase shift is imposed between the two FIDTs. The dynamic spectral broadening $2\Delta E$ was extracted from the data by fitting it to a time-integrated oscillating Lorentzian emission line \cite{Weiss2018a}. In \cref{fig:Fig2}(d), $2\Delta E$ is plotted as a function of $\Delta \phi$. Its trend follows the theoretical expression $2\Delta E = 2\Delta E_0 \times 2\vert\cos\big((\Delta \phi+\gamma)/2\big)\vert$, where $2\Delta E_0$ is the energy broadening when only one of the FIDTs is excited. The fitting parameter $\gamma=\SI{-2.0}{^\circ}$ represents a residual phase shift attributed to a slight length mismatch of the RF cables within the cryostat. The good fitting also confirms that heating has no noticeable effect, even when both FIDTs are driven simultaneously on the same chip. 

\begin{figure*}[t!]
\includegraphics[width=0.9\textwidth]{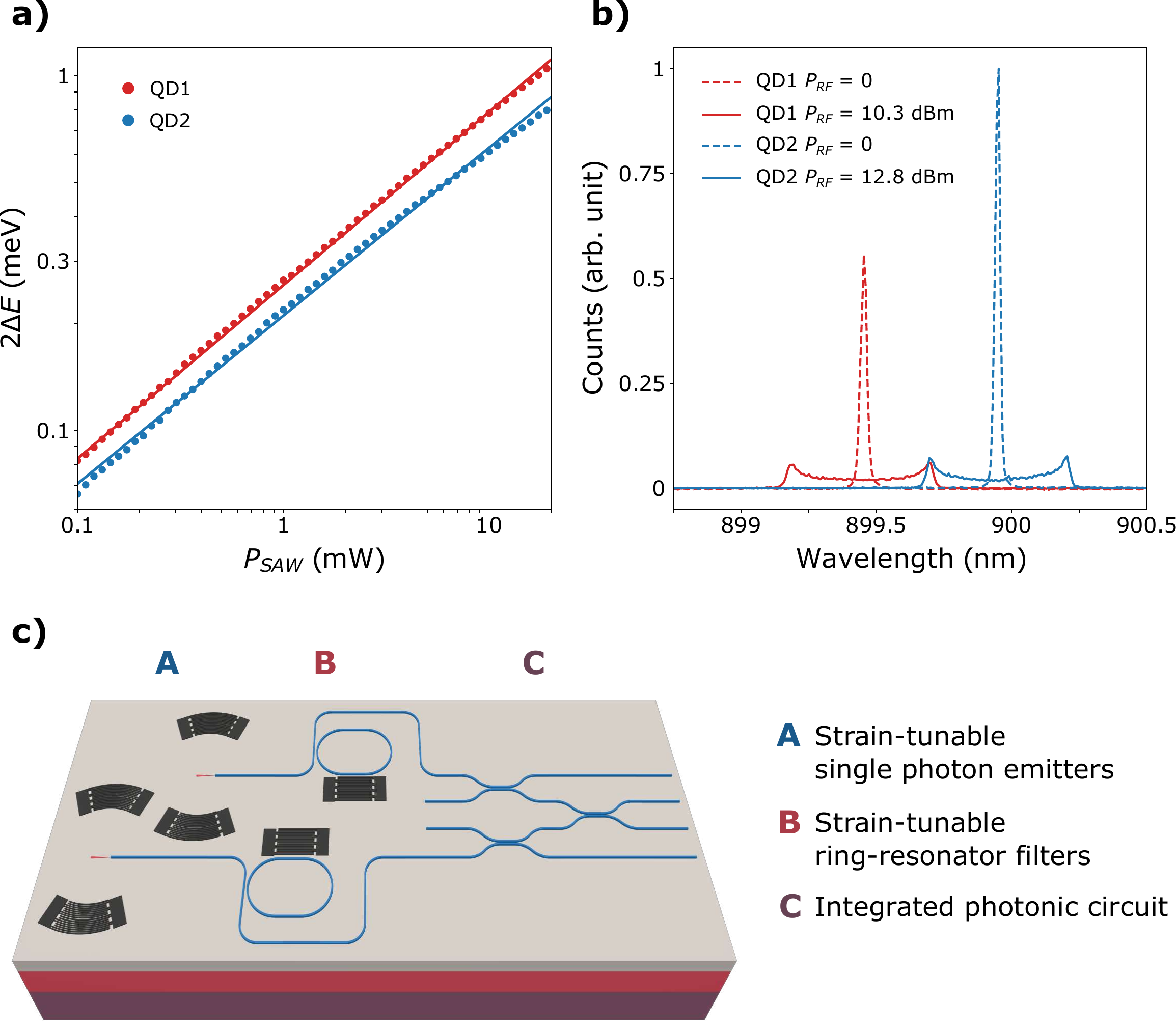}
\caption{\label{fig:Fig3}\textbf{a)} Strain-induced energy splitting of the charged exciton line of quantum dots 1 and 2 as a function of $P_{RF}$. The solid lines are linear fits. \textbf{b)} Emission peaks of QD1 and QD2 without (dashed lines) and with (solid lines) SAW-induced modulation. For all the measurements, the QDs were excited with a $\SI{80}{MHz}$ laser at $\SI{500}{nW}$. \textbf{c)} Artistic image of two strain-tunable NWQDs integrated in a hybrid thin-film LN photonic platform which comprises strain-tunable ring-resonator filters injecting resonant photons into an integrated photonic circuit.}
\end{figure*}

Similarly to QD1, the modulation performance of QD2 in the second delay line was investigated. For both QDs, the spectral broadening is plotted as a function of the driving RF power on a logarithmic scale (\cref{fig:Fig3}(a)). Over the studied power range, the modulation of NW2 is, on average, $\SI{21}{\%}$ smaller than that of NW1. This discrepancy is attributed to variations in the performance of the FIDTs, and to different adhesions of the NWs on the lithium niobate. In both cases, the strain-induced broadenings follow the power law $2\Delta E \propto (P_{\text{RF}})^{\alpha}$, where $\alpha=0.489 \pm 0.001$ for NW1, and $\alpha=0.474 \pm 0.002$ for NW2. These coefficients closely approach the ideal value of $\alpha = 0.5$ expected for deformation potential coupling, indicating that the observed broadening primarily arises from optomechanical coupling \cite{Weiss2018a}. As shown in \cref{fig:Fig3}(b), the wavelength of the charged exciton line of NW2 is $\SI{0.5}{nm}$ greater than that of NW1. The different emission wavelengths can stem from multiple factors, from the growth process \cite{Laferriere2022} to the static strain and charge environment after transfer to the host substrate. Two separate RF signals at $\SI{400}{MHz}$ were employed to excite one FIDT from each delay line in order to modulate both QDs independently. A higher power was applied to the FIDT of the second delay line to achieve the modulation amplitude necessary to achieve identical modulation amplitudes for both nanowires. Over an acoustic cycle, the two QDs emitted at a common wavelength of $\SI{899.70}{nm}$.

To extract the modulated photons from both QDs at this common wavelength, synchronized microwave sources with a $\pi$-shift between them are required, ensuring that one photon is blueshifted while the other is redshifted. 
Depending on the ratio of the exciton lifetime to the SAW period, two different processing schemes can be considered. If the SAW period is longer than the radiative decay time, the strain field and the resulting deformation potential around the QD can be considered quasi-static. Having the SAW frequency be an integer multiple of the repetition rate of the pulsed lasers allows for repeated optical excitation of the QD at a fixed point in the acoustic cycle. As a result, the emission would consistently fall within a desired energy range, eliminating the need for spectral filtering.
Conversely, if the strain field varies during the exciton recombination time, different emission wavelengths arise and post-emission filtering becomes necessary to ensure spectral overlap. This can be realized with integrated photonic resonators, as illustrated in \cref{fig:Fig3}(c), which can be tuned using electro-optic schemes \cite{Jiang2022, Wang2018c} or SAWs \cite{Shao2019b, Tadesse2014, Yu2020a}, provided a tunable phase to compensate for propagation delay.

\section{Discussion}
A statistical analysis of similar NWQDs, emitting at slightly longer wavelengths than those investigated here, revealed a Gaussian distribution of the emission wavelengths with a standard deviation of $\SI{4.65}{nm}$ \cite{Laferriere2022}. Although the measurement presented above demonstrated that two selected NWQDs could be tuned in resonance, achieving larger spectral modulation would relax the selection process. One straightforward improvement would be to increase the driving RF power beyond  $\SI{13}{dBm}$, provided that sample heating does not deteriorate spectral tuning \cite{Buhler2022}. By extrapolating the power law observed in \cref{fig:Fig3}(a), we estimate that a dynamic broadening of $\SI{1.16}{nm}$ can be reached at a microwave power of $\SI{17.1}{dBm}$ with a single FIDT, potentially bringing $\SI{10}{\%}$ of such NWQDs into resonance. To reduce ohmic losses, a lower resistivity metal such as aluminium, gold or platinum \cite{Cai2019a} could be used instead of chromium for the FIDT electrodes. Placing the QD at an anti-node of a standing-wave created by driving both FIDTs of the delay line is another effective approach to improve modulation performance by a factor of two, as demonstrated in \cref{fig:Fig2}(d). A similar effect can be obtained by positioning the QD between two SAW mirrors and exciting the acoustic cavity with only one IDT \cite{Imany2022}, thereby reducing the thermal load by half. Furthermore, the SH0 mode profile (\cref{fig:Fig1}(d)) shows that the SAW is confined in both the LN and silica layers, hence reducing the acoustic energy at the surface. Higher mechanical confinement, and thus enhanced optomechanical modulation, could be achieved by releasing the LN \cite{Vogele2020}, although this would involve a more challenging fabrication process and result in a more fragile device.
Our strain-modulation scheme can also be scaled to more than two emitters on the same chip, without additional fabrication complexity. In this regard, the footprint of the FIDT can be shrinked from a focal length of $\SI{400}{\micro m}$ to $\SI{100}{\micro m}$ with a slight reduction of the maximum transverse displacement  of the SAW by $\SI{15}{\%}$ (section S3 of the Supporting Information).

\section{Conclusion}
We successfully transferred \ce{InAsP}/\ce{InP} nanowire quantum dots on a thin-film lithium niobate platform, and heterogeneously integrated them into hybrid photonic waveguides through \ce{Si_3N_4} strip loading. By operating a single focused interdigital transducer at $\SI{400}{MHz}$, we excited and coupled a shear SAW to the energy levels of a QD, resulting in a  modulation of the emission wavelength by $\SI{0.70}{nm}$ at $\SI{13}{dBm}$. By driving both FIDTs of the delay line, we could either double this modulation or suppress it altogether, depending on the phase difference between the driving RF signals. This local and scalable strain tuning approach allowed us to bring two waveguide-integrated NWQDs with a $\SI{0.5}{nm}$ wavelength difference into resonance. This represents a crucial step towards generating indistinguishable single photons from multiple remote emitters on a single photonic chip. Photons brought into resonance can then be filtered using resonators operating at the same frequency as the FIDTs, and subsequently manipulated with photonic circuits for integrated quantum photonic applications. 

\begin{suppinfo}
Device fabrication; Photoluminescence spectrum of NW2; FIDT acoustic field simulations. 
\end{suppinfo}

\section{Author contributions}
T.D. and T.S. contributed equally to this work. \\
T.D. and T.S. fabricated the samples, performed the measurements and the simulations, and analyzed the data. P.J.P. and D.D. grew the nanowire quantum dots. All authors contributed to discussion of the results. T.D. and T.S. wrote the manuscript with inputs from all authors. T.D. conceived the experiment. T.D. and V.Z. supervised the project. 

\section{Acknowledgement}
The work was partially supported by the Knut and Alice Wallenberg (KAW) Foundation through the Wallenberg Centre for Quantum Technology (WACQT). The authors also acknowledge the support from the European Union’s Horizon 2020 Research and Innovation Programme through the project aCryComm, FET Open Grant Agreement no. 899558.

\subsection{Funding sources}
The work was partially supported by the Knut and Alice Wallenberg (KAW) Foundation through the Wallenberg Centre for Quantum Technology (WACQT). The authors also acknowledge the support from the European Union’s Horizon 2020 Research and Innovation Programme through the project aCryComm, FET Open Grant Agreement no. 899558.\\


\bibliography{references}

\clearpage
\setcounter{figure}{0}
\renewcommand{\thefigure}{S\arabic{figure}}

\begin{center}
\textbf{\LARGE Supporting Information}
\end{center}

\section{S1. Device fabrication}

\begin{figure*}[h]
\includegraphics[width=\textwidth]{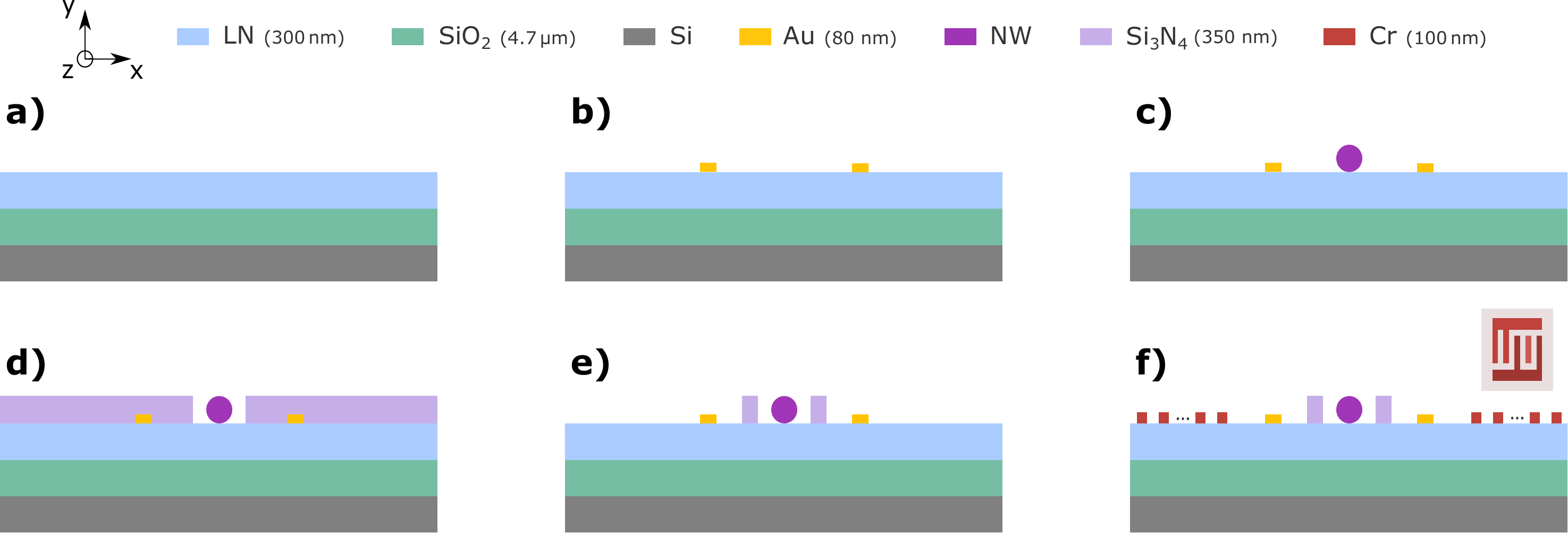}
\caption{\label{fig:FigS1}Fabrication workflow. 
}
\end{figure*}

The LNOI surface (\cref{fig:FigS1}(a)) was coated with positive resist (AR-P 6200.9), and alignment markers were patterned using electron-beam lithography (EBL). After development, a Ti/Au layer was evaporated and subsequently lifted-off (\cref{fig:FigS1}(b)). The nanowires were transferred from the growth substrate to the chip (\cref{fig:FigS1}(c)) using nano-manipulators mounted inside a scanning electron microscope (SEM). A $\SI{350}{nm}$-thick \ce{Si_3N_4} loading layer was then deposited at $\SI{300}{\celsius}$ using plasma-enhanced chemical vapor deposition (PECVD) on the entire surface (\cref{fig:FigS1}(d)). This process was carried out at \SI{1000}{mTorr} with a gas mixture of \SI{350}{sccm} \SI{5}{\%}-diluted \ce{SiH_4} and \SI{20}{sccm} \ce{NH_3}. The deposition involved repeated cycles of high-frequency plasma (\SI{13.56}{MHz} - \SI{50}{W}) and low-frequency plasma (\SI{100}{kHz} - \SI{50}{W}) for \SI{12}{s} and \SI{8}{s}, respectively. The surface was then coated with negative EBL resit (ma-N 2403) and the photonic elements were patterned by EBL according to the positions of the nanowires. The pattern was transferred to the \ce{Si_3N_4} by reactive ion etching in a \ce{CHF_3}/\ce{SF_6} plasma to define the photonic elements (\cref{fig:FigS1}(e)). The waveguides were $\SI{1.2}{\micro m}$-wide and the grating couplers had a period of \SI{590}{nm} with a filling factor of 0.5. Finally, the FIDTs were created by EBL followed by chromium evaporation and lift-off (\cref{fig:FigS1}(f)). The FIDT had a split-52 design (period showed in inset of \cref{fig:FigS1}(f)) with an electrode width of $\SI{1}{\micro m}$, allowing for SAW excitation at a fundamental frequency of $f_1=\SI{402.4}{MHz}$ and harmonics $f_n=nf_1$ for $n=2,3,$ and 4.

\section{S2. Photoluminescence spectrum of NW2}

\begin{figure*}[h]
\includegraphics[width=0.7\textwidth]{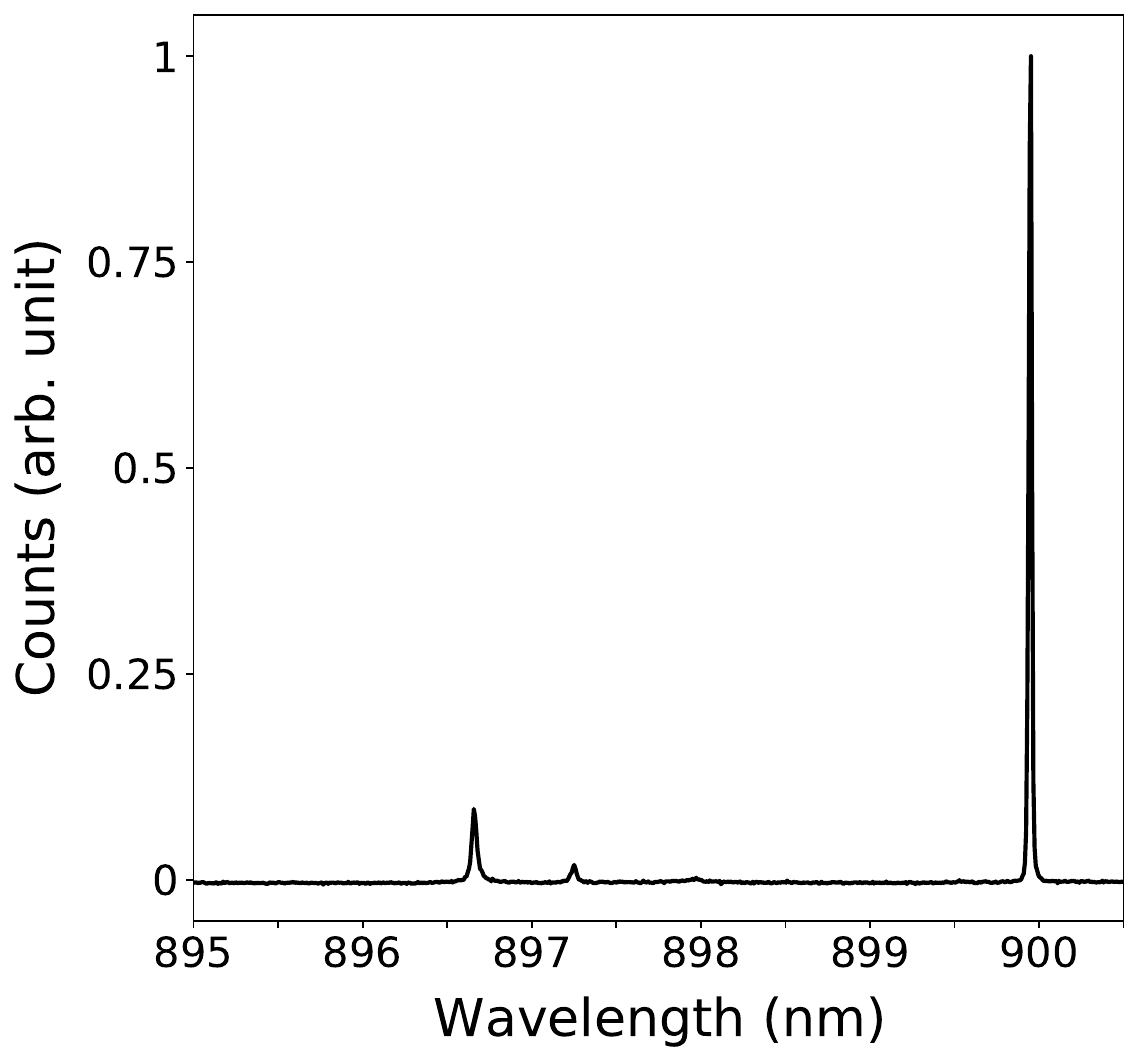}
\caption{\label{fig:FigS2}PL spectrum of QD2 without modulation. The QD was excited with a $\SI{80}{MHz}$ laser at $\SI{500}{nW}$.
}
\end{figure*}

\section{S3. FIDT acoustic field simulations}
The acoustic field generated by the FIDT is simulated using COMSOL. Twenty pairs of chromium electrodes, shaped as arcs of concentric circles, were placed on top of a $\SI{300}{nm}$-thick Y-cut thin-film LN chip with $\SI{4.7}{\micro m}$ buried oxide. The FIDT had a period of $\SI{10}{\micro m}$ with two electrodes per period, a $\SI{400}{\micro m}$ focal length and a $\SI{45}{^\circ}$ opening angle. An oscillating electric potential at $\SI{400}{MHz}$ was applied to every other electrode while the remaining electrodes were grounded. Perfectly matched layer conditions were imposed on the lateral boundaries of the domain, and the bottom boundary was fixed. The orientation of the axes is the same as shown in Fig. 1 in the main text. 

\begin{figure*}[ht]
\includegraphics[width=0.85\textwidth]{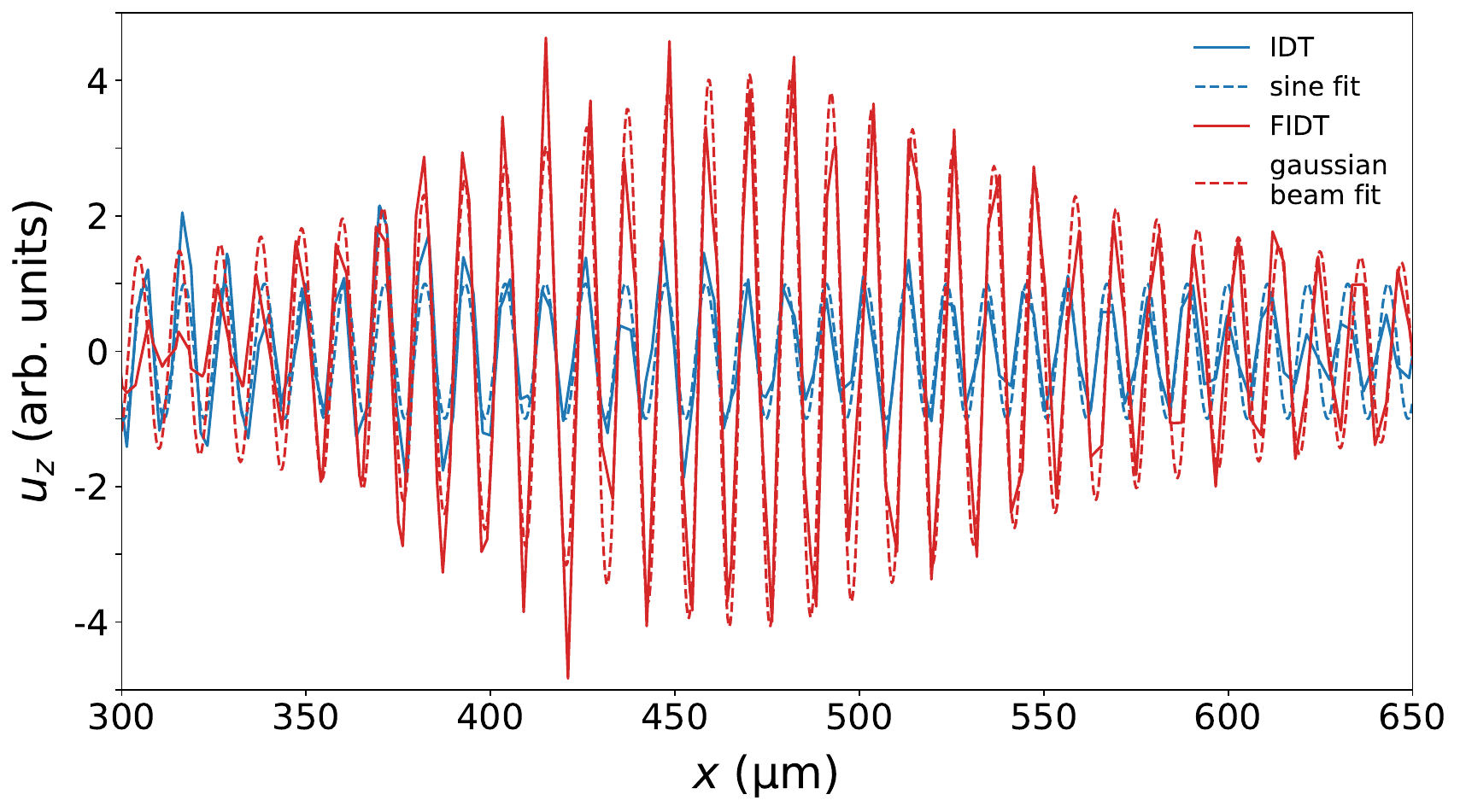}
\caption{\label{fig:FigS3}Simulated transverse displacements $u_z$ of a shear SAW propagating along the X axis of a Y-cut thin-film lithium niobate. The SAW was excited at $\SI{400}{MHz}$ by an IDT (blue) and an FIDT (red) with similar geometries. The data (solid lines) was fitted (dashed lines) to a sine function for the IDT, and to a Gaussian profile for the FIDT. 
}
\end{figure*}

\cref{fig:FigS3} shows the transverse displacement $u_z$ on the top surface along the direction of SAW propagation at a constant $z=\SI{0}{\micro m}$. The envelope of the mechanical oscillations can be fitted to a Gaussian beam profile along its center axis
\begin{equation}
    \Tilde{u}_z(x) \propto \dfrac{1}{\sqrt{1+\big((x-x_0)/x_\text{R}\big)^2}}
    \nonumber
\end{equation}
where $x_0$ is the position of the beam waist, and $x_\text{R}$ is the Rayleigh length. The fitting parameters are $x_\text{R}=\SI{60}{\micro m}$ and $x_0=\SI{470}{\micro m}$, indicating that the beam's focus is offset from the geometric focus by $\SI{70}{\micro m}$.
A similar simulation was conducted for a straight-electrode IDT with an identical period. In this case, the mechanical oscillations exhibit a nearly constant amplitude over the simulated propagation distance. This amplitude is extracted by fitting the data to a simple sinusoidal function, serving as a baseline to evaluate the performance gain of the FIDT. Compared to the IDT, the FIDT generates an acoustic field at the beam waist that is greater by a factor of 4.1, and at the geometric focal point by a factor of 2.7.

\begin{figure*}[ht]
\includegraphics[width=0.85\textwidth]{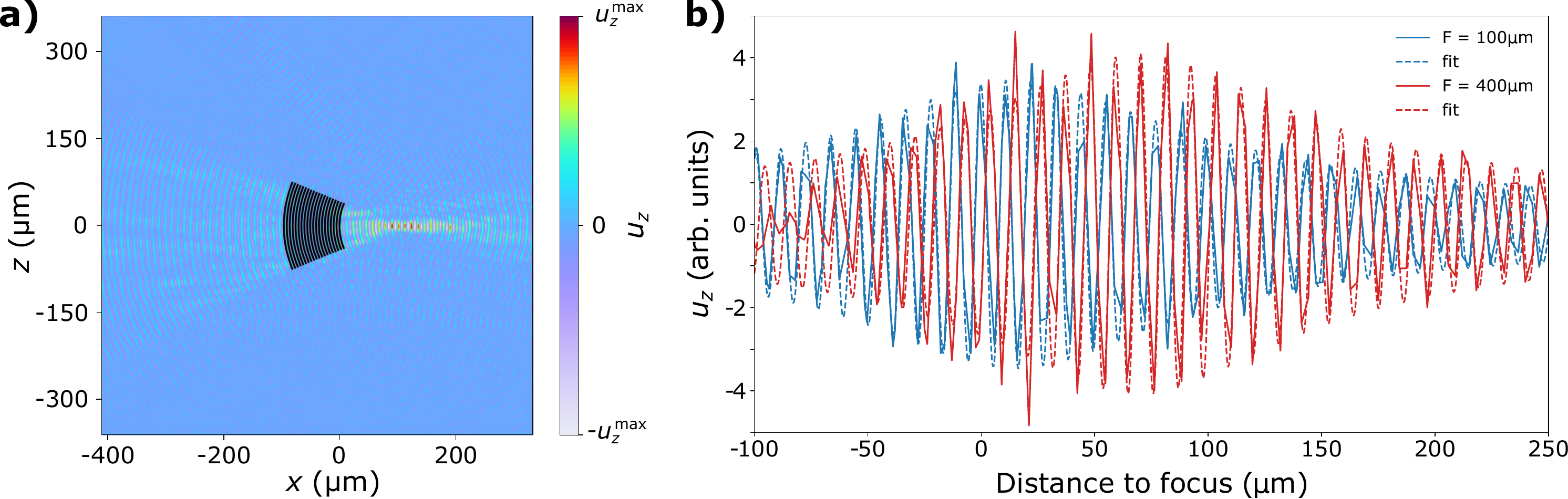}
\caption{\label{fig:FigS4}\textbf{a)}Displacement field of a SAW excited at $\SI{400}{MHz}$ by an FIDT with a $\SI{100}{\micro m}$ focal length and a $\SI{45}{^\circ}$ opening angle. The FIDT has a period of $\SI{10}{\micro m}$ with two electrodes per period. \textbf{b)} Simulated transverse displacements $u_z$ of a shear SAW propagating along the X axis of a Y-cut thin-film lithium niobate for two FIDTs with focal lengths $F$ of $\SI{400}{\micro m}$ (red) and $\SI{100}{\micro m}$ (blue). The data (solid lines) was fitted (dashed lines) to a Gaussian profiles. 
}
\end{figure*}

The transverse displacement generated by a smaller FIDT with $\SI{100}{\micro m}$ focal length but with the same $\SI{45}{^\circ}$ opening angle is shown in \cref{fig:FigS4}(a). The SAW is also focused, and the maximum displacement occurs at $\SI{15}{\micro m}$ to the geometric focus. The reduction of the footprint of the FIDT is particularly interesting for increasing the density of modulated sources on the same chip. By fitting the envelope of the displacement around the beam waist, we found that the maximum displacement generated by the $\SI{100}{\micro m}$ focal length FIDT is slightly reduced by $\SI{15}{\%}$ compared to the $\SI{400}{\micro m}$ focal length FIDT (\cref{fig:FigS4}(b)). 

\end{document}